\documentclass{elsart}

\usepackage{graphicx}

\usepackage{amssymb}

\begin{document}

\begin{frontmatter}



\title{Collective Chaos Induced by Structures of Complex Networks}

\author[label1]{Huijie Yang\corauthref{cor1}}
\author[label2]{, Fangcui Zhao}
\author[label1]{and Binghong Wang}
\address[label1]{Department of Modern Physics, University of Science and Technology of China, Hefei Anhui 230026, China}
\address[label2]{College of Life Science and Biomedical Engineering, Beijing University of Technology, Beijing 100022, China}
\corauth[cor1]{Corresponding author. E-mail: huijieyangn@eyou.com}
\begin{abstract}

Mapping a complex network of $N$coupled identical oscillators to a
quantum system, the nearest neighbor level spacing (NNLS)
distribution is used to identify collective chaos in the
corresponding classical dynamics on the complex network. The
classical dynamics on an Erdos-Renyi network with the wiring
probability $p_{ER} \le \frac{1}{N}$ is in the state of collective
order, while that on an Erdos-Renyi network with $p_{ER} >
\frac{1}{N}$ in the state of collective chaos. The dynamics on a
WS Small-world complex network evolves from collective order to
collective chaos rapidly in the region of the rewiring probability
$p_r \in [0.0,0.1]$, and then keeps chaotic up to $p_r = 1.0$. The
dynamics on a Growing Random Network (GRN) is in a special state
deviates from order significantly in a way opposite to that on WS
small-world networks. Each network can be measured by a couple
values of two parameters $(\beta ,\eta )$.
\end{abstract}

\begin{keyword}
 Complex networks \sep Collective chaos \sep Spectra statistics
\PACS 05.50.+q \sep 05.10.-a \sep 05.40.-a \sep 87.18.Sn
\end{keyword}
\end{frontmatter}

\section{Introduction}
\label{}
  Impacts of network structures on the dynamical processes attract
special attentions in recent years, to cite examples, the epidemic
spreading on networks [1-3], the response of complex networks to
stimuli [4,5] and the synchronization of dynamical systems on
networks [6-9].In this paper, we will consider the collective
motions on complex networks, just like the phonon in regular
lattices.

By means of the random matrix theory (RMT), we find that the
nearest neighbor level spacing (NNLS) distributions for spectra of
complex networks generated with WS Small-world, Erdos-Renyi and
Growing Randomly Network models can be described with Brody
distribution in a unified way. This unified description can be
used as a new measurement to characterize complex networks. The
results tell us that the topological structure of a complex
network can induce a special kind of collective motions. Under
environmental perturbations, the collective motion modes can
transition between each other abruptly. This kind of sensitivity
to outside perturbations is called collective chaos in this paper.
It is found that the properties of the collective chaos are
determined only by the structures of the networks. Without the aid
of the dynamical model presented in references [4,5], we show for
the first time that the dynamics on complex networks can be in
collective order, soft chaotic or even hard chaotic states.

\section{The Method}
\label{}

Wigner, Dyson, Mehta and others developed the Random Matrix Theory
(RMT) to understand the energy levels of complex nuclei and other
kinds of complex quantum systems [10-13]. In recent literature,
the spectral density function and the time series analysis methods
are used to capture properties of complex networks [14-19]. One of
the most important concepts in RMT is the nearest neighbor level
spacing (NNLS) distribution. A general picture emerging from
experiments and theories is that if the classical motion of a
dynamical system is regular, the NNLS distribution of the
corresponding quantum system behaves according to a Poisson
distribution. If the corresponding classical motion is chaotic,
the NNLS distribution will behave in accordance with the
Wigner--Dyson ensembles. This is the content of the famous Bohigas
conjecture [20,21]. Hence, the NNLS distribution of a quantum
system can tell us the dynamical properties of the corresponding
classical system.

Consider an undirected network of $N$coupled identical
oscillators. The Hamiltonian reads,

\begin{equation}
\label{eq1} H = \sum\limits_{n = 1}^N {h_0 (x_n ,p_n )} +
\frac{1}{2} \cdot \sum\limits_{m \ne n}^N {A_{mn} \cdot V(x_m ,x_n
)} ,
\end{equation}

\noindent where, $h_0 (x_n ,p_n )$ is the Hamiltonian of the
$n$'th oscillator, $V(x_m ,x_n )$ the coupling potential between
the $m$'th and the $n$'th oscillators and $A$ the adjacent matrix
of the network. The Hamiltonian of the corresponding quantum
system can be represented as,

\begin{equation}
\label{eq2} \hat {H} = \sum\limits_{n = 1}^N {\hat {h}_0 (x_n ,p_n
)} + \frac{1}{2} \cdot \sum\limits_{m \ne n}^N {A_{mn} \cdot \hat
{V}(x_m ,x_n )} \quad .
\end{equation}

Assuming the site energy of each oscillator is $\varepsilon _0 $
and the eigenfunction is $\varphi _0 $, we have
$\mathord{\buildrel{\lower3pt\hbox{$\scriptscriptstyle\frown$}}\over
{h}} _0 (x_n ,p_n )\varphi _0 (x_n ) = \varepsilon _0 \varphi _0
(x_n )$. The matrix elements of
$\mathord{\buildrel{\lower3pt\hbox{$\scriptscriptstyle\frown$}}\over
{H}} $ reads,

\begin{equation}
\label{eq3}
\begin{array}{l}
 H_{mn} \\
 = \left\langle {\varphi _0 (x_m )} \right|h_0 (x_m )\left| {\varphi _0 (x_n
)} \right\rangle + A_{mn} \cdot \left\langle {\varphi _0 (x_m )}
\right|V(x_m ,x_n )\left| {\varphi _0 (x_n )} \right\rangle \\
 = \varepsilon _0 \cdot \delta _{mn} + A_{mn} \cdot V_{mn} \\
 \end{array}
\end{equation}

The pattern of the spectrum of
$\mathord{\buildrel{\lower3pt\hbox{$\scriptscriptstyle\frown$}}\over
{H}} $ does not dependent on the values of $\varepsilon _0 $ and
$V_{mn} $. Assigning $\varepsilon _0 = 0$ and $V_{mn} = 1$, we
have $H = A$. By this way, the spectrum of the adjacent matrix $A$
can be used to calculate the NNLS distribution of the quantum
system. To make our discussion as self-contained as possible, we
review briefly the procedure to obtain the NNLS distribution from
the spectrum of $A$, denoted with $\left\{ {E_i \left| {i = 1,2,3,
\cdots ,N} \right.} \right\}$. $N$ is the total number of the
energy levels.

To ensure that the distances between the energy levels are
expressed in units of local mean energy level spacing, we should
first map the energy levels $E_i $ to new variables called
``unfolded energy levels'' $\xi _i $. This procedure is called
unfolding, which is generally a non-trivial task [22].

Define the cumulative density function as,

\begin{equation}
\label{eq4} G(E) \equiv N\int_{ - \infty }^E {g(s)ds} ,
\end{equation}

\noindent where $g(s)$ is the density function of the initial
energy level spectrum. We have,

\begin{equation}
\label{eq5} G(E)\left| {E_{k + 1} > E \ge E_k } \right. = k.
\end{equation}

Preprocess the spectrum, $\left\{ {E_k \left| {k = 1,2,3, \cdots
,N} \right.} \right\}$, and the corresponding accumulative density
function, $\left\{ {G(E_k )\left| {k = 1,2,3, \cdots N} \right.}
\right\}$, so that for each of them the mean is set to zero and
the variance equals to $1$, i.e.,

\begin{equation}
\label{eq6} \lambda _k = \frac{E_k - \frac{1}{N} \sum\nolimits_{m
= 1}^N {E_m } }{\left[ {\sum\nolimits_{j = 1}^N {\left( {E_j -
\frac{1}{N}\sum\nolimits_{m = 1}^N {E_m } }
\right)^2}}\right]^{\raise0.7ex\hbox{$1$} \!\mathord{\left/
{\vphantom {1
2}}\right.\kern-\nulldelimiterspace}\!\lower0.7ex\hbox{$2$}}},
\end{equation}
\begin{equation}
\label{eq7}
 F(\lambda _k ) = \frac{G(E_k ) - \textstyle{1 \over N}\sum\nolimits_{m =
1}^N {G(E_m )} }{\left[ {\sum\nolimits_{j = 1}^N {\left( {G(E_j )
- \textstyle{1 \over N}\sum\nolimits_{m = 1}^N {G(E_m )} }
\right)^2}} \right]^{\raise0.7ex\hbox{$1$} \!\mathord{\left/
{\vphantom {1
2}}\right.\kern-\nulldelimiterspace}\!\lower0.7ex\hbox{$2$}}}.
\end{equation}

Dividing the accumulative function $F(\lambda )_{ }$ into two
components, i.e., the smooth term $F_{av} (\lambda )$ and the
fluctuation term $F_f (\lambda )$, the unfolded energy levels can
be obtained as,

\begin{equation}
\label{eq8} \xi _k = F_{av} (\lambda _k ).
\end{equation}

Because we have not enough information on the accumulative density
function at present time, a polynomial is employed to describe the
relation between $\xi _k $ and $\lambda _k $, as follows,

\begin{equation}
\label{eq9} \xi _k = \sum\limits_{l = 0}^L {c_l \cdot (\lambda _k
)^l} = F_{av} (\lambda _k ).
\end{equation}
It is found that a large value of $L > 9$ can lead to a
considerable good fitting result. To guarantee the fitting results
exact enough, we assign the value as $L = 17$.

Defining the nearest neighbor level spacing (NNLS) as,

\begin{equation}
\label{eq10} \left\{ {s_i = w \cdot \frac{(\xi _{i + 1} - \xi _i
)}{\sigma _\xi }} \right\}\left| {i = 1,2,3, \cdots (N - 1)}
\right.,
\end{equation}

\noindent the Brody distribution of the NNLS reads,

\begin{equation}
\label{eq11} P(s) = \frac{\beta }{\eta }s^{\beta - 1}\exp \left[ {
- \left( {\frac{s}{\eta }} \right)^\beta } \right],
\end{equation}

\noindent which is also called Weibull distribution in the
research field of life data analysis [23]. In the definition of
NNLS,$w$ is a factor to make the values of the NNLS in a
conventional range to get a reliable fitting result, and $\sigma
_\xi = \sqrt {\frac{\sum\nolimits_{i = 1}^{N - 1} {\xi _i ^2} }{N
- 1}} $. Introducing the function, $Q(s) = \int_0^s {P(t)dt} $,
some trivial computation lead to [23],

\begin{equation}
\label{eq12} \ln R(s) \equiv \ln \left[ {\ln \left( {\frac{1}{1 -
Q(s)}} \right)} \right] = \beta \ln s - \beta \ln \eta ,
\end{equation}

\noindent based upon which we can get reliable values of the
parameters $\beta $ and $\eta $.

To obtain the function $Q(s)$, we should divide the interval where
the NNLS distributes into many bins. The size of a bin can be
chosen to be a fraction of the square root of the variance of the
NNLS, which reads, $\varepsilon = \frac{1}{R}\sqrt
{\frac{\sum\nolimits_{i = 1}^{N - 1} {s_i ^2} }{N - 1}} $. If $R$
is unreasonable small, $Q(s)$ cannot capture the exact features in
actual probability distribution function (PDF), while a much large
$R$ will induce strong fluctuations. The value of the parameter
$R$ is assigned 20 in this paper, because the fitting results are
stable in a considerable wide range about this value.

A Brody distribution reveals that the corresponding classical
complex system is in a soft chaotic state [10]. For the two
extreme conditions of $\beta = 1$ and $\beta = 2$, the Brody
distribution will reduce to the Poisson and the Wigner
distributions. And the corresponding classical complex systems are
in the hard chaotic and order states, respectively.

For the quantum system considered in this paper, it can be
initially in a quantum state of $\left| {E_n } \right\rangle $,
corresponding to the eigenvalue of $E_n $. Under a weak
environmental perturbation, the state will display completely
different behaviors [24-26]. If the system is in an order state,
the transition probability of the initial state to a new state
$\left| {E_m } \right\rangle $ will decrease rapidly with the
increase of $\left| {E_n - E_m } \right|$, and the transition
occurs mainly between the initial state and its neighboring
states. If the system is in a chaotic state, the transitions
between all the states in the chaotic regime the initial state
belongs to can occur with almost same probabilities. In the
classical dynamics, the corresponding states are the collective
motion modes, just like phonon in regular lattices. Under
perturbations the state of a chaotic system can transition between
the collective modes in the same chaotic regime abruptly, while
the state of an order system can transition between the initial
mode and its neighboring states only. Consequently, the chaotic
state of the $N$ identical oscillators is a kind of collective
behavior rather than the individual properties of each oscillator.
It is called collective chaos in this paper. The above discussions
tell us that this collective chaos dependents only on the
structure of the undirected network.

\section{Results}
\label{}

Given $N$ nodes, an Erdos-Renyi network can be constructed just by
connecting each pair with a probability $p_{ER}$ [27,28]. It is
demonstrated that there exists a critical point $p_c =
\frac{1}{N}$. For $p_{ER} < p_c $, the adjacency matrix can reduce
into many sub-matrices, the couplings between the energy levels
will be very weak and the NNLS will obey a Poisson form. For
$p_{ER} \ge p_c $, the fraction of the nodes forming the largest
sub-graph grows rapidly. The couplings between the energy levels
will become stronger and stronger, and the NNLS should obey a
Brody or even a Wigner form. Simulation results presented in Fig.1
to Fig.3 are consistent with this theoretical prediction.

\begin{figure}
\scalebox{1}[1]{\includegraphics{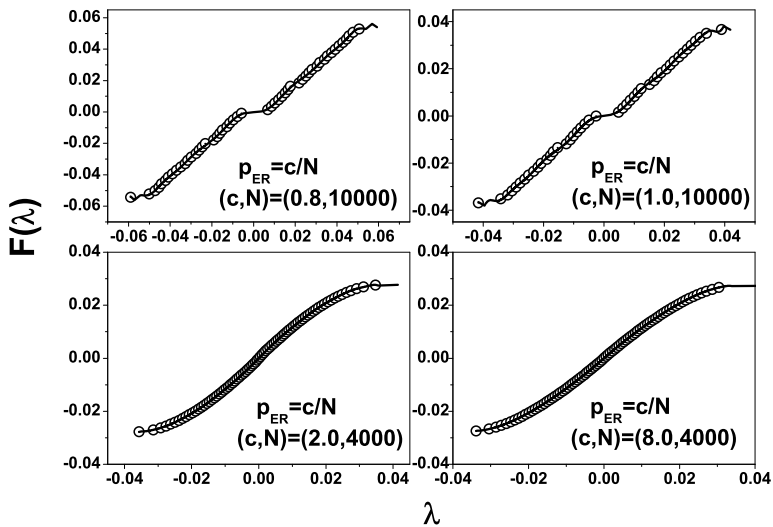}}
\caption{\label{fig:epsart} The cumulative density function of the
spectra of four Erdos-Renyi networks. The circles are the actual
values, while the solid lines are fitting results with a
17-ordered polynomial function. }
\end{figure}

\begin{figure}
\scalebox{1}[1]{\includegraphics{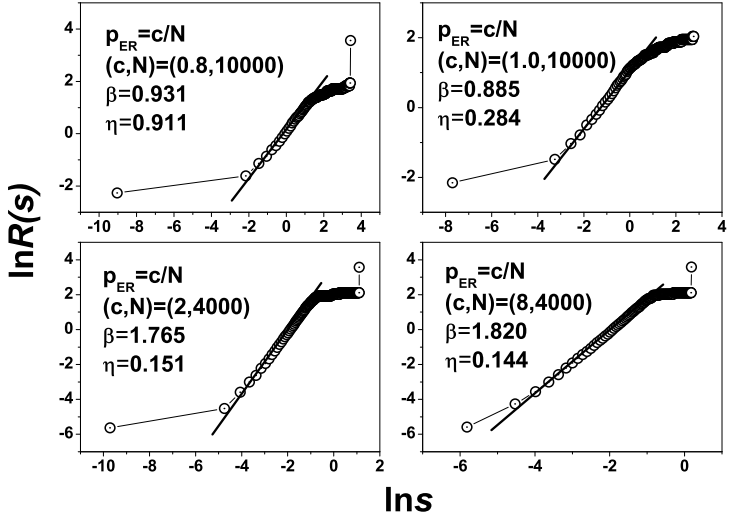}}
\caption{\label{fig:epsart} Determine the values of parameters
$(\beta,\eta)$ for the four Erdos-Renyi networks by means of the
relation presented in $Eq.12$. In the main region we are
interested the NNLS distributions obey a Brody form almost
exactly. For $p_{ER}<p_c=\frac{1}{N}$,we have $\beta=0.931 \sim
1.0$, i.e., the distribution obeys a Poisson form. For
$p_{ER}>p_c$, the distributions obey a Brody distribution form
very near the Wigner one. The errors of the parameters
$(\beta,\eta)$ are less than $0.01$.}
\end{figure}

\begin{figure}
\scalebox{1}[1]{\includegraphics{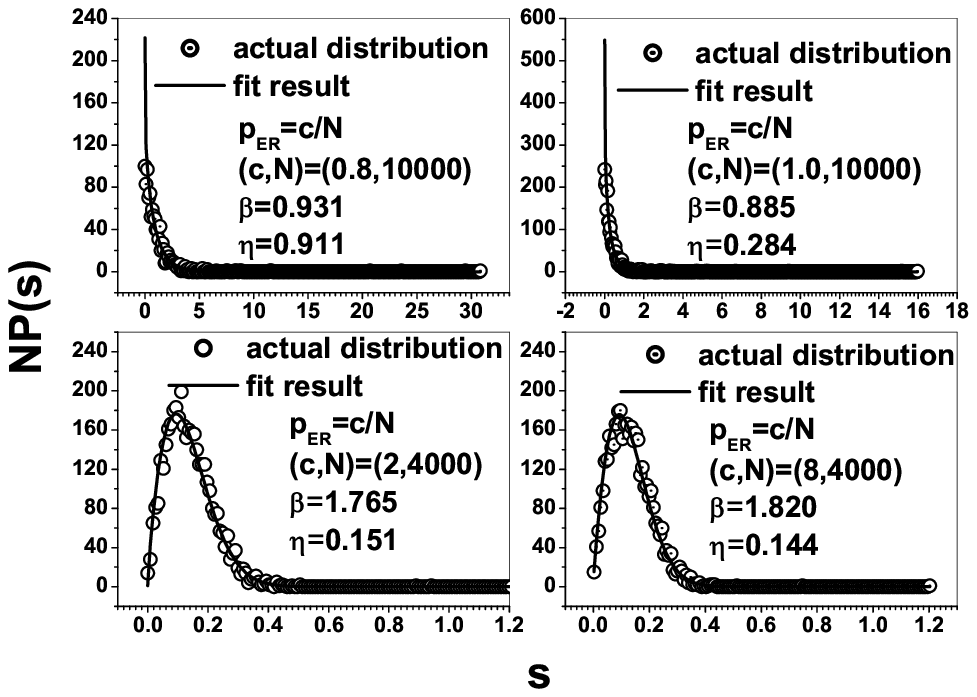}}
\caption{\label{fig:epsart} The NNLS distributions for the four
Erdos-Renyi networks. In the main regions we are interested, the
theoretical results can fit with the actual ones very well. The
errors of the parameters $(\beta,\eta)$ are less than $0.01$.}
\end{figure}

\begin{figure}
\scalebox{1}[1]{\includegraphics{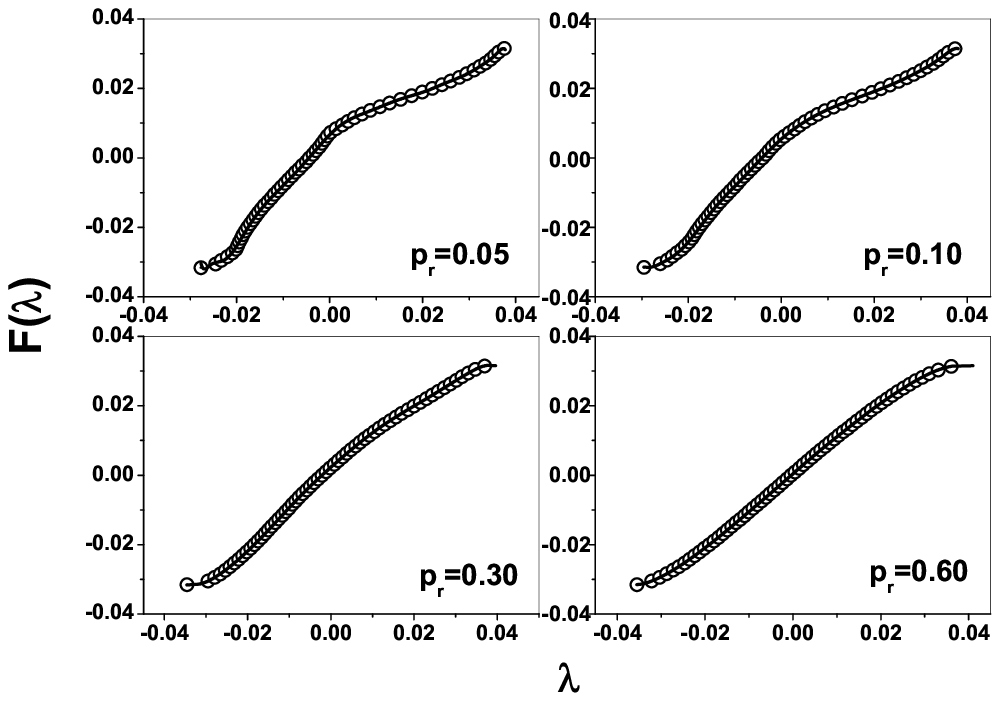}}
\caption{\label{fig:epsart} The cumulative density function of the
spectra of four selected WS small-world networks. The circles are
the actual values, while the solid lines are fitting results with
a 17-ordered polynomial function.}
\end{figure}

\begin{figure}
\scalebox{1}[1]{\includegraphics{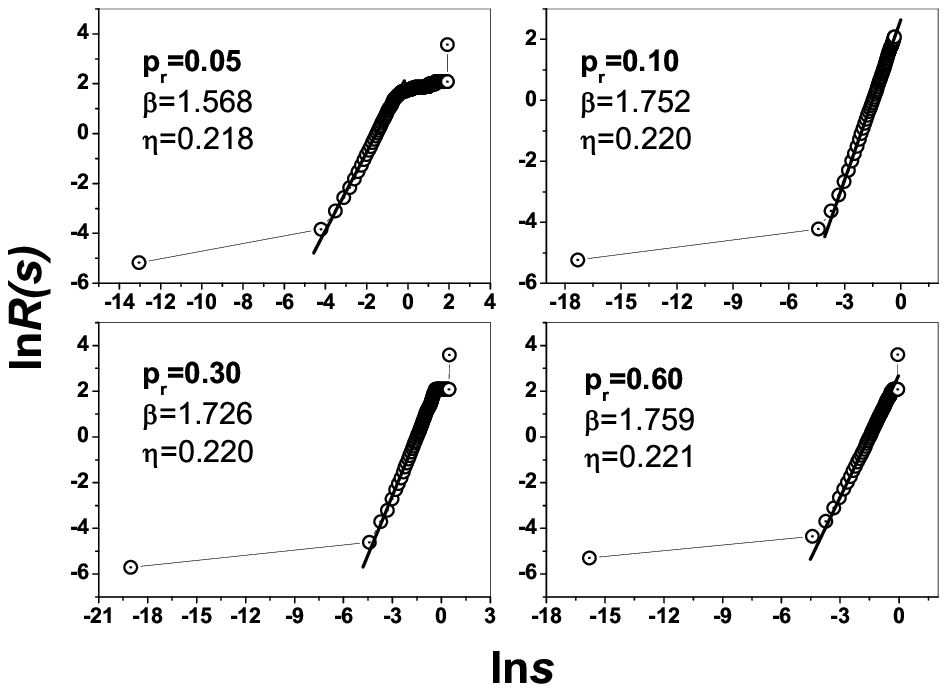}}
\caption{\label{fig:epsart} Determine the values of parameters
$(\beta,\eta)$ for the four selected WS small-world networks by
means of the relation presented in $Eq.12$. In the main region we
are interested the NNLS distributions obey a Brody form almost
exactly. The errors of the parameters $(\beta,\eta)$ are less than
$0.01$.}
\end{figure}

\begin{figure}
\scalebox{1}[1]{\includegraphics{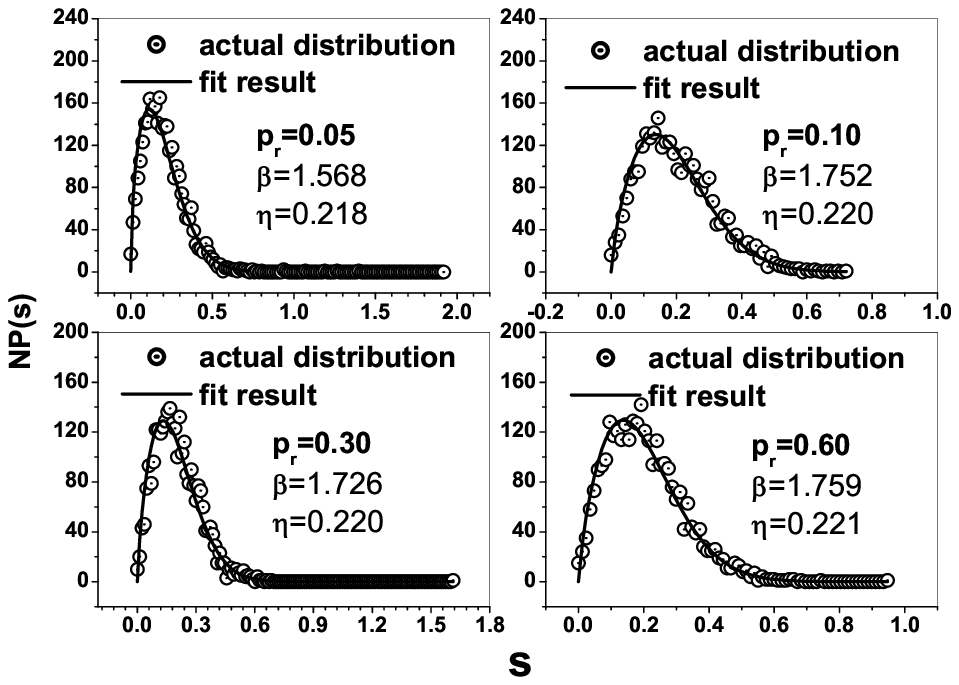}}
\caption{\label{fig:epsart} The NNLS distributions for the four
selected WS small-world networks. In the main regions we are
interested, the theoretical results can fit with the actual ones
very well. The errors of the parameters $(\beta,\eta)$ are less
than $0.01$.}
\end{figure}

\begin{figure}
\scalebox{1}[1]{\includegraphics{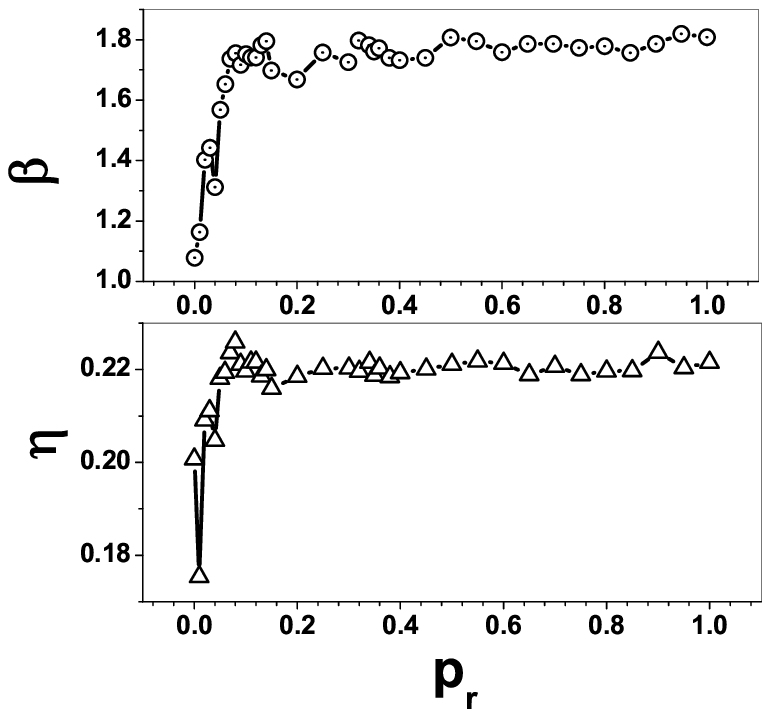}}
\caption{\label{fig:epsart} The parameters $(\beta,\eta)$ for all
the WS small-world networks constructed in this paper. In the
short region $p_r \in [0.0,0.1]$, the value of $\beta$  increases
from $1.0$ to $1.75$, i.e., the NNLS distribution evolves from a
Poisson to a near Wigner form. In the other region $p_r \in
[0.1,1]$, the networks behave almost same. Comparison tells us
that the Erdos-Renyi networks with $p_{ER}=\frac{2J}{N}(J \ge 1)$
are similar with the WS small-world network with $p_r=1$. The
errors of the parameters are less than  $0.01$.}
\end{figure}

Secondly, we consider the one-dimensional lattice small-world
model designed by Watts and Strogatz (WS small-world model) [29].
Take a one-dimensional lattice of nodes with periodic boundary
conditions, and join each node with its $k$ right-handed nearest
neighbors. Going through each edge in turn and with probability
$p_r $ rewiring one end of this edge to a new node chosen
randomly. During the rewiring procedure double edges and
self-edges are forbidden. Numerical simulations by Watts and
Strogatz show that this rewiring process allows the small-world
model to interpolate between a regular lattice and a random graph
with the constraint that the minimum degree of each node is fixed
[29]. The parameter $k$ is chosen to be 2, and $N$ is 3000. Fig.4
to Fig.6 present some typical results for different values of
rewiring probability. In the main region we are interested, the
NNLS distribution can be described with a Brody distribution
almost exactly. Fig.7 presents the values of the parameters $\beta
$ and $\eta $ versus the rewiring probability $p_r $. In a short
region $p_r \in [0.0,0.1]$, the value of $\beta $ increases from
$1$ to $1.75$, i.e., the NNLS distribution evolves from a Poisson
to a near Wigner form. In the other region $p_r \in [0.1,1.0]$ the
networks behave almost same. Comparison tells us that the
Erdos-Renyi networks with $p_{ER} = \frac{2J}{N}(J \ge 1)$ are
similar with the WS small-world network with $p_r = 1$.

\begin{figure}
\scalebox{1}[1]{\includegraphics{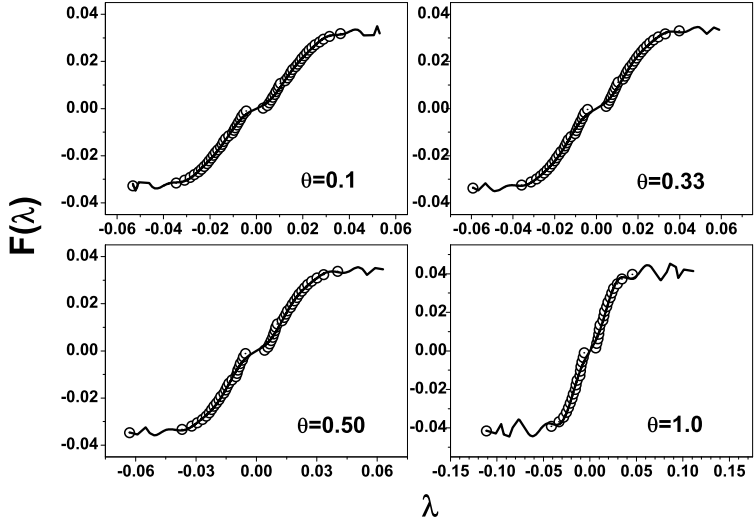}}
\caption{\label{fig:epsart} The cumulative density function of the
spectra of four selected GRN networks. The circles are the actual
values, while the solid lines are fitting results with a
17-ordered polynomial function.}
\end{figure}

\begin{figure}
\scalebox{1}[1]{\includegraphics{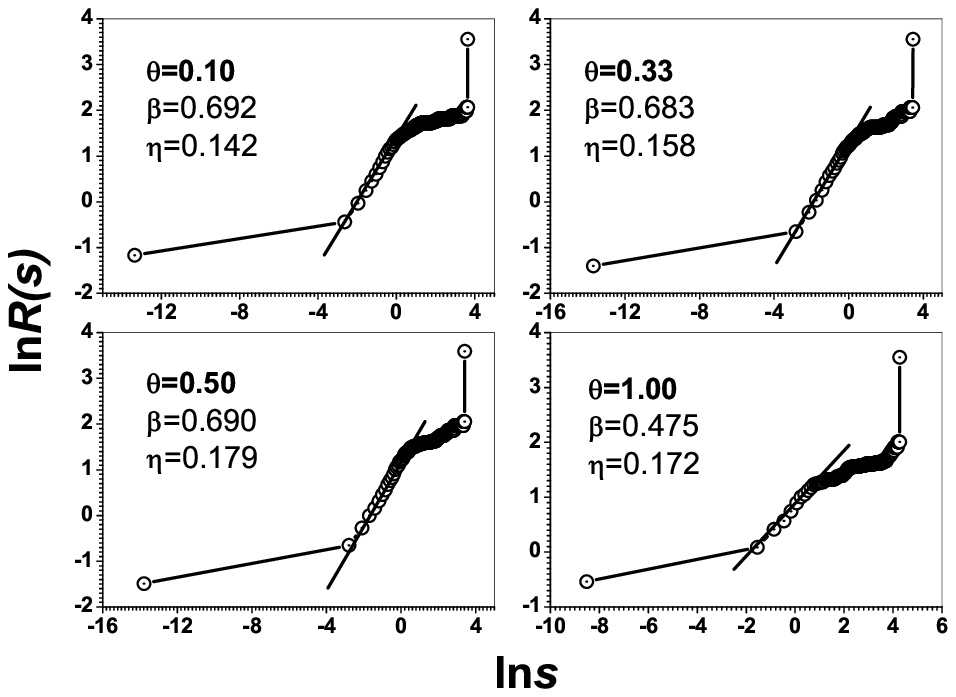}}
\caption{\label{fig:epsart} Determine the values of parameters
$(\beta,\eta)$ for the four selected GRN networks by means of the
relation presented in $Eq.12$. In the main region we are
interested the NNLS distributions obey a Brody form almost
exactly. The errors of the parameters $(\beta,\eta)$ are less than
$0.01$.}
\end{figure}

\begin{figure}
\scalebox{1}[1]{\includegraphics{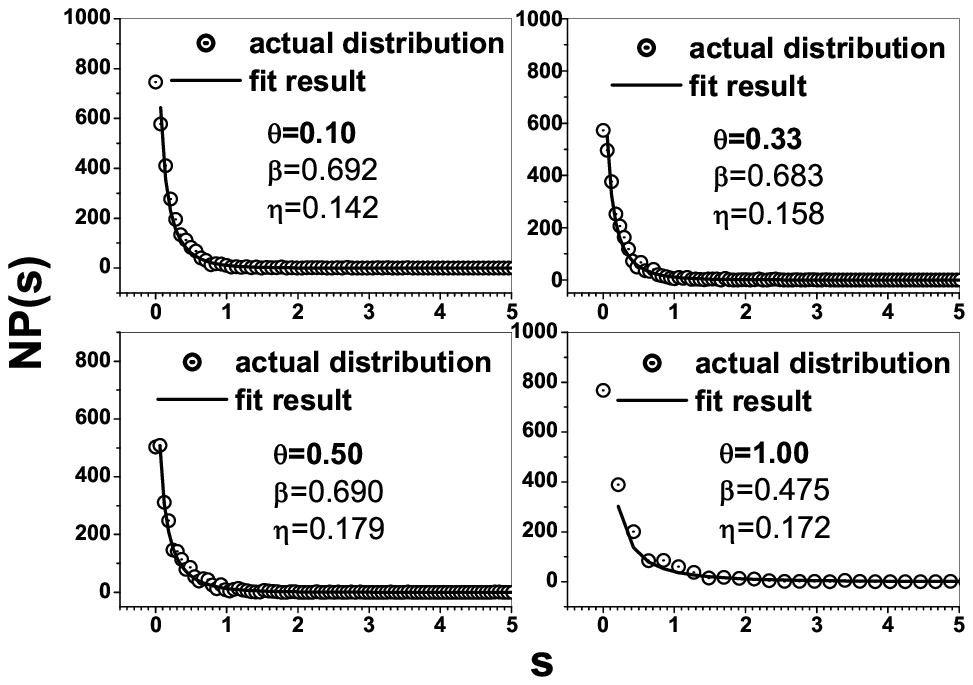}}
\caption{\label{fig:epsart} The NNLS distributions for the four
selected GRN networks. In the main regions we are interested, the
theoretical results can fit with the actual ones very well. The
errors of the parameters $(\beta,\eta)$ are less than $0.01$.}
\end{figure}

\begin{figure}
\scalebox{1}[1]{\includegraphics{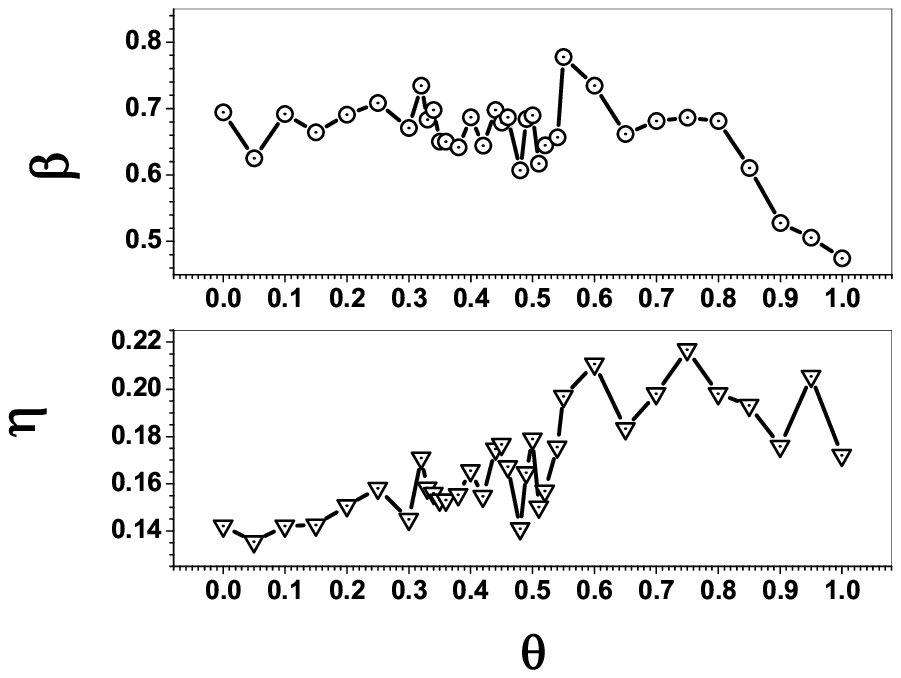}}
\caption{\label{fig:epsart} The parameters $(\beta,\eta)$ for all
the GRN networks constructed in this paper. In a large region $p_r
\in [0.0,0.8]$, the value of $\beta$ oscillates around $0.68$,
i.e, the NNLS distributions deviate significantly from the Poisson
form in a way opposite to that of WS small-world networks. In the
other region $p_r \in [0.8,1]$, the value of $\beta$ decreases
rapidly to $\sim 0.5$. The errors of the parameters $(\beta,\eta)$
are less than $0.01$.}
\end{figure}

The third considered is the growing random networks (GRN) model
[30]. Giving several connected seeds, at each step a new node is
added and a link to one of the earlier nodes is created. The
connection kernel $A_k $, defined as the probability a newly
introduced node links to a preexisting node with $k$ degree,
determines the structure of this network. A group of GRN networks
determined by a special kind of kernel, $A_k \propto k^\theta (0
\le \theta \le 1)$, are considered in this present paper. For this
kind of networks, the degree distributions decrease as a stretched
exponential in $k$. Setting $\theta = 1$ we can obtain a
scale-free network.

Fig.8 to Fig.10 present some typical results for GRN networks.
Fig.11 shows that in a wide range of $0 \le \theta \le 0.8$, the
value of the parameter $\beta $ oscillates basically around
$0.68$, i.e., the NNLS distributions deviate significantly from
the Poisson form in a way opposite to that of WS small-world
networks. In the other region $p_r \in [0.8,1.0]$ the value of
$\beta $ decreases rapidly to $\sim 0.50$. The values of the
parameter $\eta $ are also presented.

Fig.12 shows the relation between $\beta $ and $\eta $. Each point
represents a complex network. The results for the three kinds of
networks are all illustrated. The WS small-world networks and the
GRN networks are separated by the Poisson form, i.e., $\beta = 1$,
significantly. The Erdos-Renyi networks with $p_{ER} \le p_c =
\textstyle{1 \over N}$ obey near Poisson distribution, while that
with $p_{ER} > p_c = \textstyle{1 \over N}$ are similar with the
almost complete random WS small-world networks ($p_r \sim 1.0)$.

\begin{figure}
\scalebox{1}[1]{\includegraphics{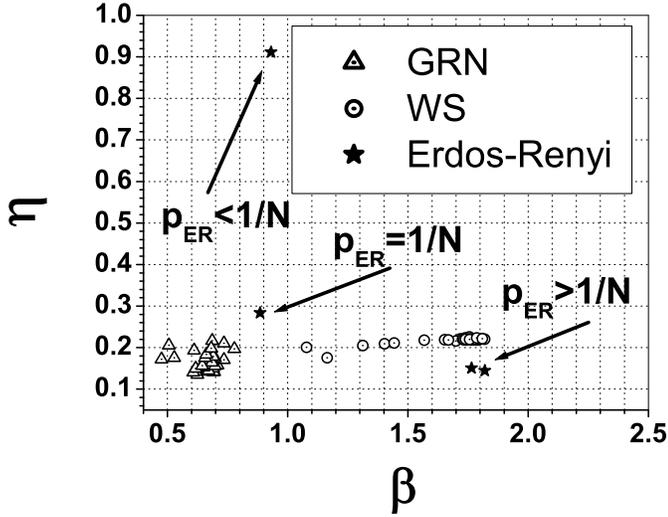}}
\caption{\label{fig:epsart} The relation between the two
parameters $(\beta,\eta)$ . Each point corresponds to a complex
network. The results for the three kinds of networks are all
illustrated. The WS small-world networks and the GRN networks are
separated by the Poisson form, i.e., $\beta=1$ , significantly.
The Erdos-Renyi networks with $p_{ER} <p_c=\frac{1}{N}$  obey near
Poisson distribution, while that with $p_{ER} \ge p_c$  are
similar with the almost complete random WS small-world networks
($p_r \sim 1$ ). The position of a network in this scheme may tell
some useful information. The errors of the parameters
$(\beta,\eta)$ are less than $0.01$.}
\end{figure}

\section{Summary}
\label{} In summary, based upon the RMT theory we investigate the
NNLS distributions for the ER, the WS Small-world and the GRN
networks. The Brody form can describe all these distributions in a
unified way. The NNLS distributions of the quantum systems of the
network of $N$ coupled identical oscillators tell us that the
corresponding classical dynamics on the Erdos-Renyi networks with
$p_{ER} < p_c = \frac{1}{N}$ are in the state of collective order,
while that on the Erdos-Renyi networks with $p_{ER} > p_c =
\frac{1}{N}$ in the state of collective chaos. On WS small-world
networks, the classical dynamics evolves from collective order to
collective chaos rapidly in the region $p_r \in [0.0,0.1]$, and
then keeps chaotic up to $p_r = 1.0$. For GRN model, contrary to
that on the WS small world networks, the classical dynamics are in
special states deviate from order significantly in an opposite
way.

These dynamical characteristics are determined only by the
structures of the considered complex networks. In a very recent
paper [31], the authors point out that for some biological
networks the NNLS distributions obey the Wigner form. The dynamics
on these networks should be in a state of collective chaos. And
the removal of nodes may change this dynamical characteristic from
collective chaos to collective order. Therefore, constructing a
mini network with selected key nodes should be considered
carefully in discussing the collective dynamics on a complex
network.

Without the aid of the simplified model of dynamics presented in
[3,4] we obtain the dynamical characteristics on complex networks.
The NNLS distribution can capture directly the relation between
the structure of a complex network and the dynamics on it.

It should be pointed out that the collective chaos induced by the
structures of complex networks is completely different with the
individual chaotic states of the oscillators on the networks. For
a network with regular structure, the classical dynamical
processes on it should display collective order, even if the
coupled identical oscillators on the nodes may be in chaotic
states. On the contrary, for a complex network with a complex
structure (e.g., a WS small-world structure), the classical
dynamical processes on it should display collective chaos, even if
the coupled identical oscillators on the nodes may be in order
states as harmonic oscillations.

Each network can be measured by a couple values of $(\beta ,\eta
)$. The position of a complex network in $\beta $ versus $\eta $
scheme may tell us useful information for classification of real
world complex networks.

\section{Acknowledgements}
\label{}
This work was partially supported by the National Natural
Science Foundation of China under Grant No. 70471033, 10472116 and
No.70271070. One of the authors (H. Yang) would like to thank
Prof. Y. Zhuo, Prof. J. Gu in China Institute of Atomic Energy and
Prof. S. Yan in Beijing Normal University for stimulating
discussions.

\end{document}